\documentclass[12,preprint]{aastex}

\begin{document}

\title{Afterglow Light Curves from Impulsive Relativistic Jets with
an Unconventional Structure}

\author{Jonathan Granot}

\affil{Kavli Institute for Particle Astrophysics and Cosmology,
Stanford University, P.O. Box 20450, MS 29, Stanford, CA 94309;
granot@slac.stanford.edu}

\begin{abstract}
  
  The jet structure in gamma-ray burst (GRB) sources is still largely
  an open question. The leading models invoke either (i) a roughly
  uniform jet with sharp edges, or (ii) a jet with a narrow core and
  wide wings where the energy per solid angle drops as a power law
  with the angle $\theta$ from the jet symmetry axis. Recently, a two
  component jet model has also been considered, with a narrow uniform
  jet of initial Lorentz factor $\Gamma_0\gtrsim 100$ surrounded by a
  wider uniform jet with $\Gamma_0\sim 10-30$. Some models predict
  more exotic jet profiles, such as a thin uniform ring (i.e. the
  outflow is bounded by two concentric cones of half opening angle
  $\theta_c$ and $\theta_c+\Delta\theta$, with
  $\Delta\theta\ll\theta_c$) or a fan (a thin outflow with
  $\Delta\theta\ll 1$ along the rotational equator,
  $\theta_c=\pi/2+\Delta\theta/2$).  In this paper we calculate the
  expected afterglow light curves from such jet structures, using a
  simple formalism that is developed here for this purpose, and could
  also have other applications. These light curves are qualitatively
  compared to observations of GRB afterglows. It is shown that the two
  component jet model cannot produce sharp features in the afterglow
  model due to the deceleration of the wide jet or the narrow jet
  becoming visible at lines of sight outside of it. We find that a
  ``ring" shaped jet or a ``fan" shaped jet produce a jet break in the
  afterglow light curve that is too shallow compared to observations,
  where the change in the temporal decay index across the jet break is
  about half of that for a uniform conical jet.  For a ``ring" jet,
  the jet break is divided into two distinct and smaller breaks, the
  first occurring when $\gamma\Delta\theta\sim 1-2$ and the second
  when $\gamma\theta_c\sim 1/2$.

\end{abstract}

\keywords{gamma-rays: bursts --- gamma-rays: theory --- ISM: jets and
  outflows --- relativity --- radiation mechanisms: nonthermal}

\section{Introduction}
\label{sec:intro}

Different lines of evidence suggest that GRB outflows are collimated
into narrow jets. An indirect but compelling argument comes from the
very high values for the energy output in gamma-rays assuming
isotropic emission, $E_{\rm \gamma,iso}$, that are inferred for GRBs
with known redshifts, $z$, which approach and in one case (GRB 991023)
even exceed $M_\odot c^2$. Such extreme energies in an
ultra-relativistic outflow are hard to produce in models involving
stellar mass progenitors. If the outflow is collimated into a narrow
jet which occupies a small fraction, $f_b\ll 1$, of the total solid
angle, then the strong relativistic beaming due to the very high
initial Lorentz factor ($\Gamma_0\gtrsim 100$) causes the emitted
gamma-rays to be similarly collimated.  This reduces the true energy
output in gamma-rays by a factor of $f_b^{-1}$ to $E_\gamma=f_b
E_{\rm\gamma,iso}$, thus significantly reducing the energy
requirements. A more direct line of evidence in favor of a narrowly
collimated outflow comes from achromatic breaks seen in the afterglow
light curves of many GRBs \citep{Rhoads97,Rhoads99,SPH99}.

Since the initial discovery of GRB afterglows in the X-ray
\citep{Costa97}, optical \citep{vanParadijs97} and radio
\citep{Frail97}, many afterglows have been detected and the quality of
individual afterglow light curves has improved dramatically
\citep[e.g.,][]{Lipkin04}. Despite all the observational and theoretical
progress, the structure of GRB jets remains largely an open question.
This question is of great importance and interest, as it is related to
issues that are fundamental for our understanding of GRBs, such as their
event rate, total energy, and the requirements from the compact source
that accelerates and collimates these jets.

The leading models for the jet structure are: (i) the uniform jet (UJ)
model
\citep{Rhoads97,Rhoads99,PM99,SPH99,KP00,MSB00,Granot01,Granot02},
where the energy per solid angle, $\epsilon$, and the initial Lorentz
factor, $\Gamma_0$, are uniform within some finite half-opening angle,
$\theta_j$, and sharply drop outside of $\theta_j$, and (ii) the
universal structured jet (USJ) model \citep{LPP01,RLR02,ZM02}, where
$\epsilon$ and $\Gamma_0$ vary smoothly with the angle $\theta$ from
the jet symmetry axis. In the UJ model the different values of the jet
break time, $t_j$, in the afterglow light curve arise mainly due to
different $\theta_j$ (and to a lesser extent due to different ambient
densities). In the USJ model, all GRB jets are intrinsically
identical, and the different values of $t_j$ arise mainly due to
different viewing angles, $\theta_{\rm obs}$, from the jet
axis\footnote{In fact, the expression for $t_j$ is similar to that for
a uniform jet with $\epsilon\to\epsilon(\theta=\theta_{\rm obs})$ and
$\theta_j\to\theta_{\rm obs}$.}. The observed correlation, $t_j\propto
E_{\rm \gamma,iso}^{-1}$ \citep{Frail01,BFK03}, implies a roughly
constant true energy, $E$, between different GRB jets in the UJ model,
and $\epsilon\propto\theta^{-2}$ outside of some core angle,
$\theta_c$, in the USJ model \citep{RLR02,ZM02}. This is assuming a
constant efficiency, $\epsilon_\gamma$, for producing the observed
prompt $\gamma$-ray or X-ray emission. If the efficiency
$\epsilon_\gamma$ depends on the $\theta$ in the USJ model, for
example, then different power laws of $\epsilon$ with $\theta$ are
possible \citep{GGB05}, such as a core with wings where
$\epsilon\propto\theta^{-3}$, as is obtained in simulations of the
collapsar model \citep{ZWM03,ZWH04}.

Other jet structures have also been proposed in the literature. A jet
with a Gaussian angular profile \citep{KG03,ZM02} may be thought of as
a more realistic version of a uniform jet, where the edges are smooth
rather than sharp. If both $\epsilon(\theta)$ and $\Gamma_0(\theta)$
have a Gaussian profile (corresponding to a constant rest mass per
solid angle in the outflow) then the afterglow light curves are rather
similar to those for a uniform jet \citep{KG03}. If, on the other
hand, $\epsilon(\theta)$ is Gaussian while\footnote{This corresponds
to a Gaussian angular distribution of the rest mass per solid angle,
i.e. very little mass near the outer edge of the jet, which is the
opposite of what might be expected due to mixing near the walls of the
funnel in the massive star progenitor.} $\Gamma_0(\theta)={\rm
const}$, then the light curves for off-axis viewing angles
(i.e. outside the core of the jet) have a much higher flux at early
times, compared to a Gaussian $\Gamma_0(\theta)$ or a uniform jet, due
to a dominant contribution from the emitting material along the line
of sight \citep{GR-RP05}. Such a jet structure was considered as a
quasi-universal jet model \citep{Zhang04}.

Another jet structure that received some attention recently is a
two-component jet model \citep{Pedersen98,Frail00,
Berger03,Huang04,PKG05,WDHL05} with a narrow uniform jet of initial
Lorentz factor $\Gamma_0\gtrsim 100$ surrounded by a wider uniform jet
with $\Gamma_0\sim 10-30$. Theoretical motivation for such a jet
structure has been found both in the context of the cocoon in the
collapsar model \citep{R-RCR02} and in the context of a
hydromangetically driven neutron rich jet \citep{VPK03}. The light
curves for this jet structure have been calculated analytically
\citep{PKG05} or semi-analytically \citep{Huang04,WDHL05}, and it has
been suggested that this model can account for sharp bumps (i.e. fast
rebrightening episodes) in the afterglow light curves of GRB 030329
\citep{Berger03} and XRF 030723 \citep{Huang04}.  Here we show that
effects such as the integration over the surface of equal arrival time
of photons to the observer and the gradual hydrodynamic transition at
the deceleration epoch smoothen the resulting features in the
afterglow light curve, so that they cannot produce features as sharp
as those mentioned above.

\begin{figure}
\includegraphics[width=15cm]{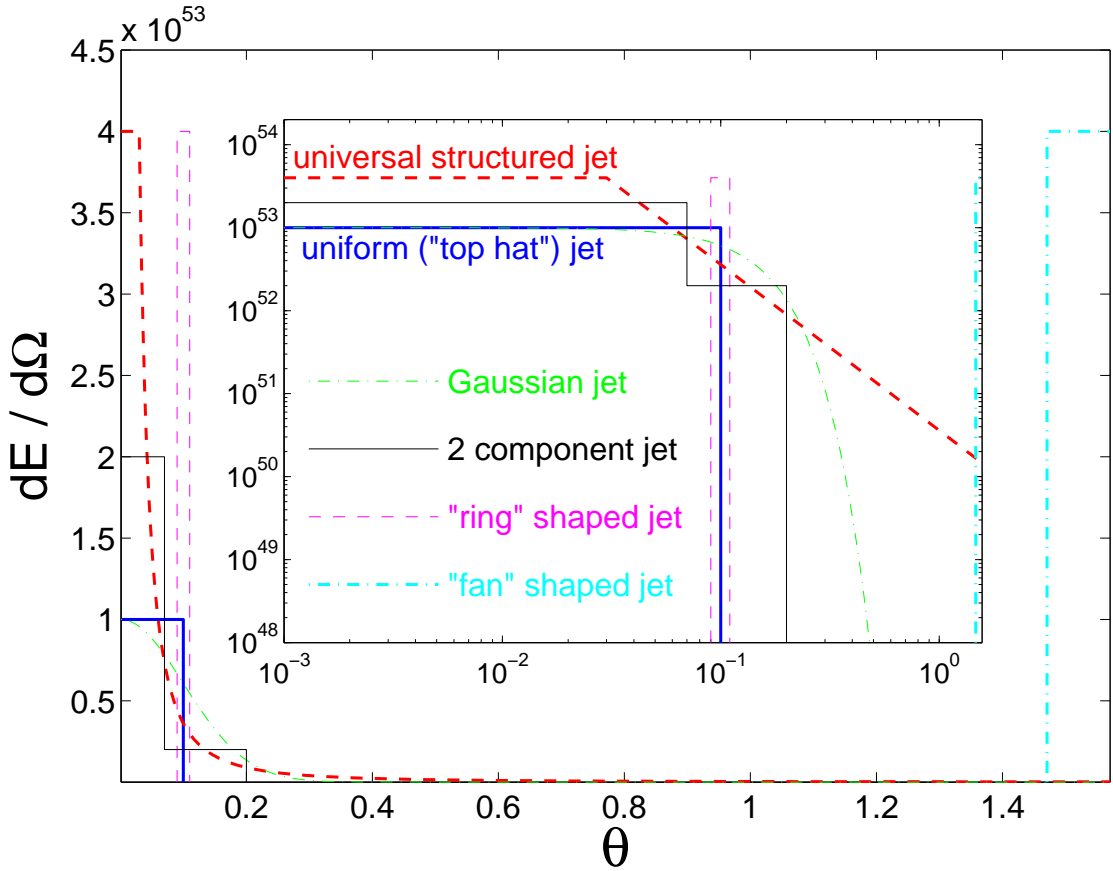}
\caption{A schematic diagram of the energy per solid angle,
  $\epsilon=dE/d\Omega$, for the various jet structures that are
  discussed in this paper, shown both in a linear scale ({\it main figure})
  and in a log-log scale ({\it big inset at the center}).}
\label{jet_structures}
\end{figure}

More ``exotic'' jet structures have also been considered. One example
is a jet with a cross section in the shape of a ``ring'', sometimes
referred to as a ``hollow cone'' \citep{EL03,LE04,EL04,LB05}, which is
uniform within $\theta_c<\theta<\theta_c+\Delta\theta$ where
$\Delta\theta\ll\theta_c$.  Another example is a ``fan'' or ``sheet''
shaped jet \citep{Thompson04} where a magneto-centrifugally launched
wind from the proto-neutron star, formed during the supernova
explosion in the massive star progenitor becomes relativistic as the
density in its immediate vicinity drops, and is envisioned to form a
thin sheath of relativistic outflow which is somehow able to penetrate
through the progenitor star along the rotational equator, forming a
relativistic outflow within $\Delta\theta\ll 1$ around $\theta=\pi/2$
(or $\theta_c=\pi/2-\Delta\theta/2$). We stress that this has been
suggested as a possible jet structure within this model, but the final
jet structure is by no means clear, and other jet structure might also
be possible within this model (T.~A. Thompson, private communication).
The various jet structures are shown schematically in
Fig. \ref{jet_structures}.

The light curves for the more conventional jet structures, namely the
UJ and USJ models as well as the Gaussian jet model, have been
calculated in detail. The light curves for the less conventional jet
structures, however, have either been calculated using a simple
analytic or semi analytic model (for the two-component jet model), or
have not been considered at all (for the ``ring'' of ``fan'' jet
structures). In this paper we calculate the afterglow light curves for
these models.  In \S \ref{sec:thin_sphere} a simple formalism is
developed for calculating the observed emission from a thin spherical
relativistic shell, which includes integration over the surface of
equal arrival time of photons to the observer. It is generalized to a
uniform ring shaped jet, where a finite range in $\theta$
($\theta_c<\theta<\theta_c+\Delta\theta$) is occupied by the outflow,
in \S \ref{sec:uniform_ring}. The final expression for the observed
flux for any viewing angle is in the form of a sum of two one
dimensional integrals, which is trivial to evaluate numerically. This
formalism is used to calculate the light curves for various jet
structures in \S \ref{sec:res}. In \S \ref{sec:2comp} it is applied to
a uniform jet and a two component jet, while in \S\S \ref{sec:ring}
and \ref{sec:fan} it is used to calculate the light curves for a ring
shaped jet and a fan shaped jet, respectively. Our conclusions are
discussed in \S \ref{sec:conc}.

\section{Calculating the Light Curve from a Relativistic Spherical thin Shell}
\label{sec:thin_sphere}

The emitting shell is assumed to be relativistic, with a Lorentz
factor $\gamma=(1-\beta^2)^{-1/2}\gg 1$, and infinitely thin, so that
its location is described by its radius $R$ as a function of the lab
frame time $t$. The thin shell approximation is valid in the limit
where the width of the shell is $\Delta R\ll R/\gamma^2$. Typically,
$\Delta R\sim R/\gamma^2$ so that the thin shell approximation is only
marginally valid, and an integration over the radial profile of the
shell might introduce some changes of order unity to the resulting
light curve \citep[e.g.,][]{GPS99,GS02}. In this work, however, we
neglect the radial structure of the emitting shell for the sake of
simplicity, and in order to stress the effects of the jet angular
structure.

A photon emitted at a radius $R$, at a lab frame time $t$, and at
an angle $\theta$ from the line of sight, reaches the observer at
an observed time
\begin{equation}\label{T}
T=t-R\cos\theta\ .
\end{equation}
The radius is given by
\begin{equation}\label{R}
R=c\int_0^t d\tilde{t}\beta(\tilde{t})\approx
ct-\int_0^R\frac{d\tilde{R}}{2\gamma^2(\tilde{R})}\ ,
\end{equation}
where the last expression holds in the relativistic limit
($\gamma\gg 1$). Let us assume a power law external density profile,
$\rho_{\rm ext}=AR^{-k}$. The deceleration radius is given by
\begin{equation}\label{R_dec}
R_{\rm dec}=\left[\frac{(3-k)E_{\rm iso}}{4\pi
A\eta^2c^2}\right]^{1/(3-k)}=
\left\{\matrix{2.5\times 10^{16}n_0^{-1/3}E_{\rm
  iso,52}^{1/3}\eta_{2.5}^{-2/3}\;{\rm cm} & \quad & k=0\ ,\cr\cr
1.8\times 10^{13}A_*^{-1}E_{\rm iso,52}\eta_{2.5}^{-2}\;{\rm cm} & \quad &
k=2\ ,}\right.
\end{equation}
where $\eta=10^{2.5}\eta_{2.5}$ is the initial Lorentz factor, $E_{\rm
iso}=10^{52}E_{\rm iso,52}$ is the isotropic equivalent energy,
$n=n_0\;{\rm cm^{-3}}$ is the external density for a uniform external
medium ($k=0$) and $A=5\times 10^{11}A_*\;{\rm g\;cm^{-1}}$ for a
stellar wind environment ($k=2$). The corresponding observed
deceleration time is
\begin{equation}\label{T_dec}
T_{\rm dec}=\frac{R_{\rm dec}}{2c\eta^2}=
\left\{\matrix{4.2(1+z)n_0^{-1/3}E_{\rm
  iso,52}^{1/3}\eta_{2.5}^{-8/3}\;{\rm s} & \quad & k=0\ ,\cr\cr
3.0\times 10^{-3}(1+z)A_*^{-1}E_{\rm iso,52}\eta_{2.5}^{-4}\;{\rm s} & \quad &
k=2\ .}\right.
\end{equation}
The Lorentz factor as a function of radius is given by
\begin{equation}\label{gamma}
\gamma(R)\approx\left\{\matrix{\eta & R<R_{\rm dec}\cr\cr
\eta(R/R_{\rm dec})^{-(3-k)/2}
& R>R_{\rm dec}}\right.\ ,
\end{equation}
\citep{BM76}.  If the bulk velocity of the emitting fluid is in the
radial direction, as we assume here, then the flux density is given
by\footnote{more generally, $\beta\cos\theta$ should be replaced by
  $\hat{n}\cdot\vec{\beta}$, where $\hat{n}$ is the direction to the
  observer (in the lab frame), and if the angle $\theta$ is not
  measured from the line of sight, then $R\cos\theta$ should be
  replaced by $\hat{n}\cdot\vec{r}$.}
\begin{eqnarray}\label{F_nu}
F_\nu(T)&=&\frac{(1+z)}{d_L^2(z)}\int
d^4x\,\,\delta\left(t-\frac{T}{1+z}-\frac{R\cos\theta}{c}\right)
\frac{j'_{\nu'}}{\gamma^2(1-\beta\cos\theta)^2}
\\ \nonumber
&=& \frac{(1+z)}{4\pi d_L^2(z)}\int
dt\,\,\delta\left(t-\frac{T}{1+z}-\frac{R\cos\theta}{c}\right)
\int\frac{dL'_{\nu'}}{\gamma^3(1-\beta\cos\theta)^3}\ ,
\end{eqnarray}
where primed quantities are measured in the local rest frame of the
emitting fluid, $j'_{\nu'}$ is the spectral emissivity (emitted energy
per unit volume, frequency, time and solid angle), $L'_{\nu'}$ is in
the spectral luminosity (the total emitted energy per unit time and
frequency, assuming a spherical emitting shell), $z$ and
$d_L(z)=10^{28}d_{L28}\;$cm are the redshift and luminosity distance
of the source, respectively, and
$\nu'=(1+z)\gamma(1-\beta\cos\theta)\nu$.  We have
$dL'_{\nu'}=L'_{\nu'}(R)d\phi d\cos\theta/4\pi$ where
$L'_{\nu'}(R)\propto R^a(\nu')^b$ and the values of the power law
indexes $a$ and $b$ depend on the power law segment of the spectrum
\citep{Sari98}, and are calculate explicitly below.

For simplicity we ignore the self absorption frequency, and assume
that the spectrum at any given time is described by three power law
segments that are divided by two break frequencies, $\nu_m$ and
$\nu_c$ \citep{SPN98}. We also consider only the emission from the
shocked external medium, and do not take into account the emission
from the reverse shock. Now, $L'_{\nu',{\rm max}}\propto B' N_e$,
where $B'\propto\gamma\rho_{\rm ext}^{1/2}$ is the magnetic field
(which is assumed to hold a constant fraction, $\epsilon_B$, of the
internal energy in the shocked matter), and $N_e\propto R^{3-k}$ is
the total number of emitting electrons behind the forward shock. Also,
$\nu'_m\propto B'\gamma_m^2\propto B' \gamma^2$ and $\nu'_c\propto
B'\gamma_c^2\propto \gamma^{-1}R^{3k/2-2}$, which imply
\begin{eqnarray}\label{L_nu_max}
L'_{\nu',{\rm max}}&\propto&\left\{\matrix{ R^{3-3k/2} & R<R_{\rm
dec}\cr R^{3/2-k} & R>R_{\rm dec}}\right.\ ,\\ \label{nu_m} \nu'_m
&\propto &\left\{\matrix{ R^{-k/2} & R<R_{\rm dec}\cr
R^{-(9-2k)/2} & R>R_{\rm dec} }\right.\ ,\\
\nu'_c &\propto &\left\{\matrix{ R^{3k/2-2} & R<R_{\rm dec}\cr
R^{k-1/2} & R>R_{\rm dec} }\right.\ .
\end{eqnarray}
For fast cooling ($\nu_c<\nu_m$) we find
\begin{eqnarray}\label{FC1}
L'_{\nu'}(R<R_{\rm dec})&\propto&\left\{\matrix{
R^{11/3-2k}(\nu')^{1/3} & \nu'<\nu'_c\cr R^{2-3k/4}(\nu')^{-1/2} &
\nu'_c<\nu'<\nu'_m\cr R^{2-k(p+2)/4}(\nu')^{-p/2} &
\nu'>\nu'_m}\right.\ ,\\
\label{FC2} L'_{\nu'}(R>R_{\rm dec})&\propto&\left\{\matrix{
R^{(5-4k)/3}(\nu')^{1/3} & \nu'<\nu'_c\cr
R^{(5-2k)/4}(\nu')^{-1/2} & \nu'_c<\nu'<\nu'_m\cr
R^{[14-9p+2k(p-2)]/4}(\nu')^{-p/2} & \nu'>\nu'_m}\right.\ ,
\end{eqnarray}
while for slow cooling ($\nu_c>\nu_m$) we have
\begin{eqnarray}\label{SC1}
L'_{\nu'}(R<R_{\rm dec})&\propto&\left\{\matrix{
R^{3-k/2}(\nu')^{1/3} & \nu'<\nu'_m\cr
R^{3-k(p+5)/4}(\nu')^{(1-p)/2} & \nu'_m<\nu'<\nu'_c\cr
R^{2-k(p+2)/4}(\nu')^{-p/2} &
\nu'>\nu'_c}\right.\ ,\\
\label{SC2} L'_{\nu'}(R>R_{\rm dec})&\propto&\left\{\matrix{
R^{3-4k/3}(\nu')^{1/3} & \nu'<\nu'_m\cr
R^{[15-9p-2k(3-p)]/4}(\nu')^{(1-p)/2} & \nu'_m<\nu'<\nu'_c\cr
R^{[14-9p+2k(p-2)]/4}(\nu')^{-p/2} & \nu'>\nu'_c}\right.\ ,
\end{eqnarray}
where $p$ is the power law index of the electron energy distribution.
The values of $a$ and $b$ that are defined by $L'_{\nu'}(R)\propto
R^a(\nu')^b$ are given in Table \ref{table1}. The value of $a$ changes
at radii corresponding to hydrodynamic transitions, such as $R_{\rm
dec}$, where the ejecta stops coasting and starts to decelerate
significantly. If there is significant lateral spreading at $R_j$ (the
radius associated with the jet break time, $T_j$, in the afterglow
light curve) then this would cause a change in the value of $a$
between $R_{\rm dec}<R<R_j$ and $R_j<R<R_{\rm NR}$. A similar change
in the value of $a$ occurs at the radius of the non-relativistic
transition, $R_{\rm NR}$. In this work, however, we concentrate on the
relativistic regime ($\gamma\gg 1$ and $R\ll R_{\rm NR}$).

\section{Calculating the Light Curves from a Jet with a Uniform Ring
  Angular Profile}
\label{sec:uniform_ring}

We now specify for a jet with an angular profile of a uniform ring,
with an inner half-opening angle $\theta_c$ and an angular width
$\Delta\theta$,
\begin{equation}\label{epsilon}
\epsilon=\frac{dE}{d\Omega}=\left\{\matrix{\epsilon_0 &
\theta_c<\tilde{\theta}<\theta_c+\Delta\theta\cr\cr 0 & {\rm
otherwise}}\right.\ ,
\end{equation}
where $\tilde{\theta}$ is the angle from the symmetry axis of the jet,
which is located at an angle $\theta_{\rm obs}$ from the line of
sight. Assuming a double sided jet, the true energy is
$E=4\pi[\cos\theta_c-\cos(\theta_c+\Delta\theta)]\epsilon_0\approx
2\pi\Delta\theta(2\theta_c+\Delta\theta)\epsilon_0\approx
4\pi\theta_c\Delta\theta\epsilon_0$, where the second (third) equality
holds in the limit $\theta_c,\Delta\theta\ll 1$
($\Delta\theta\ll\theta_c$).

We note that this jet structure can be used to describe not only a
``ring" shaped jet, but also a uniform jet with sharp edges, a two
component jet,  or a ``fan" shaped jet. A uniform jet of
half-opening angle $\theta_j$ corresponds to $\theta_c=0$ and
$\Delta\theta=\theta_j$. A two component jet with a narrow (wide)
jet component of half-opening angle $\theta_n$ ($\theta_w$)
corresponds to the sum of two rings, the first a uniform jet with
$\theta_c=0$ and $\Delta\theta=\theta_n$ and the second a ring with
$\theta_c=\theta_n$ and $\Delta\theta=\theta_w-\theta_n$. A ``fan"
shaped jet corresponds to the limit of a very thin ring,
$\Delta\theta\ll\theta_c$, as long as $\gamma\theta_c\gg 1$, i.e. as
long as the visible region of angle $\sim\gamma^{-1}$ around the
line of sight is small compared to the half-opening angle of the
ring. It can also be directly modeled by
$\theta_c=\pi/2-\Delta\theta/2$ with $\Delta\theta\ll 1$.

For simplicity, we neglect the lateral spreading of the jet. This is
also motivated by the results of numerical studies
\citep{Granot01,KG03} which show a very modest degree of lateral
expansion as long as the jet is sufficiently relativistic.

For a given observed time $T$, $R$ is a function of $\theta$ alone,
according to Eq. \ref{T}, and does not depend on the azimuthal angle
$\phi$. This also applies, within the jet itself, to all the physical
quantities in the integrand in Eq. \ref{F_nu}, that are a function of
$R$: $L'_{\nu'}(R)$, $\gamma(R)$ and $\beta(R)$. Outside of the jet,
however, there is no contribution to the flux. Therefore, we need to
determine the fraction, $\Delta\phi/2\pi$, of a circle of angle
$\theta$ from the line of sight which intersects the emitting ring,
and multiply the integrand in Eq. \ref{F_nu} by this factor.  It is
most convenient to work in spherical coordinates where the z-axis
points to the observer, and the jet axis is within the x-z plane
(i.e. at $\phi=0$). The intersection of a cone of half-opening angle
$\theta$ around the line of sight with the inner and outer edges of
the ring shaped jet occurs at $\phi_1$ and $\phi_2$, respectively,
which are given by\footnote{Let
$\hat{n}(\theta,\phi)=\hat{z}\cos\theta+\hat{y}\sin\theta\sin\phi
+\hat{x}\sin\theta\cos\phi$ be a unit vector in the direction
described by the angles $(\theta,\phi)$ in polar coordinates. The
inner and outer edges of the ring shaped jet are given by
$\cos\alpha=\hat{n}(\theta_{\rm
obs},0)\cdot\hat{n}(\theta,\phi)=\cos\theta_{\rm
obs}\cos\theta+\sin\theta_{\rm obs}\sin\theta\cos\phi$, where
$\alpha=\theta_c$ and $\theta_c+\Delta\theta$ for the inner and outer
edges of the jet, respectively. Now for a given value of $\theta$ this
gives us the value of $\phi$ at which a cone of half-opening angle
$\theta$ around the line of sight intersects the inner and outer edges
of the jet: $\cos\phi=(\cos\alpha-\cos\theta_{\rm
obs}\cos\theta)/\sin\theta_{\rm obs}\sin\theta$.}
\begin{eqnarray}\label{phi1}
\cos\phi_1 &=& \frac{\cos\theta_c-\cos\theta_{\rm
obs}\cos\theta}{\sin\theta_{\rm
obs}\sin\theta}\approx\frac{\theta_{\rm
obs}^2+\theta^2-\theta_c^2}{2\theta_{\rm obs}\theta}\ ,\\
\cos\label{phi2} \phi_2
&=&\frac{\cos(\theta_c+\Delta\theta)-\cos\theta_{\rm
obs}\cos\theta}{\sin\theta_{\rm
obs}\sin\theta}\approx\frac{\theta_{\rm
obs}^2+\theta^2-(\theta_c+\Delta\theta)^2}{2\theta_{\rm
obs}\theta}\ ,
\end{eqnarray}
where the second expression approximately hold when all relevant
angles ($\theta_c$, $\Delta\theta$, $\theta_{\rm obs}$, $\theta$)
are $\ll 1$. We find
\begin{equation}\label{dphi_0}
\frac{\Delta\phi}{2\pi}(\theta_{\rm obs}=0)=\left\{\matrix{1 &
\theta_c<\theta<\theta_c+\Delta\theta\cr\cr 0 & {\rm
otherwise}}\right.\ ,
\end{equation}
\begin{equation}\label{dphi_1}
\frac{\Delta\phi}{2\pi}\left(0<\theta_{\rm
obs}<\frac{\Delta\theta}{2}\right)=\left\{\matrix{ 0 &
\theta<\theta_c-\theta_{\rm obs}\cr\cr 1-\phi_1/\pi &
\theta_c-\theta_{\rm obs}<\theta<\theta_c+\theta_{\rm obs}\cr\cr 1 &
\theta_c+\theta_{\rm obs}<\theta<\theta_c+\Delta\theta-\theta_{\rm
obs}\cr\cr \phi_2/\pi\ \ & \theta_c+\Delta\theta-\theta_{\rm
obs}<\theta_c+\Delta\theta+\theta_{\rm obs}\cr\cr 0 &
\theta>\theta_c+\Delta\theta+\theta_{\rm obs}}\right.\ ,
\end{equation}
\begin{equation}\label{dphi_2}
\frac{\Delta\phi}{2\pi}\left(\frac{\Delta\theta}{2}<\theta_{\rm
obs}<\theta_c\right)=\left\{\matrix{ 0 & \theta<\theta_c-\theta_{\rm
obs}\cr\cr 1-\phi_1/\pi & \theta_c-\theta_{\rm
obs}<\theta<\theta_c+\Delta\theta-\theta_{\rm obs}\cr\cr
(\phi_2-\phi_1)/\pi\ \  & \theta_c+\Delta\theta-\theta_{\rm
obs}<\theta<\theta_c+\theta_{\rm obs}\cr\cr \phi_2/\pi\ \  &
\theta_c+\theta_{\rm obs}<\theta<\theta_c+\Delta\theta+\theta_{\rm
obs}\cr\cr
 0 & \theta>\theta_c+\Delta\theta+\theta_{\rm
obs}}\right.\ ,
\end{equation}
\begin{equation}\label{dphi_3}
\frac{\Delta\phi}{2\pi}\left(\theta_c<\theta_{\rm
obs}<\theta_c+\frac{\Delta\theta}{2}\right)=\left\{\matrix{ 1 &
\theta<\theta_{\rm obs}-\theta_c\cr\cr 1-\phi_1/\pi & \theta_{\rm
obs}-\theta_c<\theta<\theta_c+\Delta\theta-\theta_{\rm obs}\cr\cr
(\phi_2-\phi_1)/\pi & \theta_c+\Delta\theta-\theta_{\rm
obs}<\theta<\theta_c+\theta_{\rm obs}\cr\cr \phi_2/\pi\ \ &
\theta_c+\theta_{\rm obs}<\theta<\theta_c+\Delta\theta+\theta_{\rm
obs}\cr\cr
 0 & \theta>\theta_c+\Delta\theta+\theta_{\rm
obs}}\right.\ ,
\end{equation}
\begin{equation}\label{dphi_4}
\frac{\Delta\phi}{2\pi}\left(\theta_c+\frac{\Delta\theta}{2}<\theta_{\rm
obs}<\theta_c+\Delta\theta\right)=\left\{\matrix{ 1 &
\theta<\theta_c+\Delta\theta-\theta_{\rm obs}\cr\cr \phi_2/\pi &
\theta_c+\Delta\theta-\theta_{\rm obs}<\theta<\theta_{\rm
obs}-\theta_c\cr\cr (\phi_2-\phi_1)/\pi & \theta_{\rm
obs}-\theta_c<\theta<\theta_c+\theta_{\rm obs}\cr\cr \phi_2/\pi &
\theta_c+\theta_{\rm obs}<\theta<\theta_c+\Delta\theta+\theta_{\rm
obs}\cr\cr
 0 & \theta>\theta_c+\Delta\theta+\theta_{\rm
obs}}\right.\ ,
\end{equation}
\begin{equation}\label{dphi_5}
\frac{\Delta\phi}{2\pi}\left(\theta_{\rm
obs}>\theta_c+\Delta\theta\right)=\left\{\matrix{ 0 &
\theta<\theta_{\rm obs}-\theta_c-\Delta\theta\cr\cr \phi_2/\pi &
\theta_{\rm obs}-\theta_c-\Delta\theta<\theta<\theta_{\rm
obs}-\theta_c\cr\cr (\phi_2-\phi_1)/\pi & \theta_{\rm
obs}-\theta_c<\theta<\theta_c+\theta_{\rm obs}\cr\cr \phi_2/\pi &
\theta_c+\theta_{\rm obs}<\theta<\theta_c+\Delta\theta+\theta_{\rm
obs}\cr\cr
 0 & \theta>\theta_c+\Delta\theta+\theta_{\rm
obs}}\right.\ .
\end{equation}

It is more convenient to integrate over $R$, instead of over
$\theta$. In the relativistic limit,
\begin{equation}\label{T2}
T\approx\left\{\matrix{\frac{R}{2\eta^2c}(1+\eta^2\theta^2) &
R<R_{\rm dec}\cr\cr
\frac{R}{2\gamma^2c}\left[\frac{1}{(4-k)}+\gamma^2\theta^2\right]
+\left(\frac{3-k}{4-k}\right)\frac{R_{\rm dec}}{2\eta^2c}= & \cr
\frac{R_L}{2(4-k)\gamma_L^2c}\left[(4-k)\gamma_L^2\theta^2x
+x^{4-k}+(3-k)x_{\rm dec}^{4-k}\right] & R>R_{\rm dec} }\right.\ ,
\end{equation}
where $\gamma_L=\gamma(R_L)$ and $R_L(T)$ are the Lorentz factor
and radius from which a photon emitted along the line of sight (at
$\theta=0$) reaches the observer at an observed time $T$, while
$x\equiv R/R_L$, and $x_{\rm dec}\equiv R_{\rm dec}/R_L$. We have
\begin{eqnarray}\label{R_L}
R_L(T)&=&\left\{\matrix{2\eta^2c\,T & T<T_{\rm dec}\cr\cr
\frac{2(4-k)\gamma_L^2cT}{1+(3-k)x_{\rm dec}^{4-k}} & T>T_{\rm
dec} }\right.\ ,
\\ \nonumber
\\ \label{x_dec}
x_{\rm dec}&=& [(4-k)T/T_{\rm dec}-(3-k)]^{-1/(4-k)}\ ,
\end{eqnarray}
and thus obtain
\begin{equation}\label{theta_square}
\theta^2=\left\{\matrix{\frac{1}{\eta^2}\left(\frac{2\eta^2c\,T}{R}-1\right)
 & R<R_{\rm dec}\cr\cr
\frac{1}{(4-k)\gamma_L^2}(x^{-1}-x^{3-k})
& R>R_{\rm dec} }\right.\ .
\end{equation}
Therefore, $d\cos\theta\approx d(\theta^2)/2=dR[d(\theta^2)/dR]/2$
where
\begin{equation}\label{dcostheta_dx}
\frac{d\cos\theta}{dx}\approx\frac{R_L}{2}\frac{d(\theta^2)}{dR}=
\left\{\matrix{-(cT/R_L)x^{-2} & x<x_{\rm dec}\cr\cr
-\frac{x^{-2}+(3-k)x^{2-k}}{2(4-k)\gamma_L^2} & x>x_{\rm dec}
}\right.\ .
\end{equation}
Finally, we can express the integral for the flux density
as\footnote{For simplicity, from this point on we drop all the
cosmological corrections, and simply denote the distance to the
source by $D$ (they can easily be put back in at the final result,
according to Eq. \ref{F_nu}).}
\begin{eqnarray}\nonumber
F_\nu(T)&=&\frac{1}{4\pi D^2}
\int\frac{dL'_{\nu'}}{\gamma^3(1-\beta\cos\theta)^3}
\\ \nonumber
&=&\frac{1}{8\pi
D^2}\int_0^{R_L(T)}dR\,\left|\frac{d\cos\theta}{dR}\right|\delta^3(R)
L'_{\nu'}(R)\frac{\Delta\phi[\theta(R)]}{2\pi}
\\  \label{ F_nu} &=&\frac{1}{8\pi D^2}\int_0^1
dx\,\left|\frac{d\cos\theta}{dx}\right|\delta^3L'_{\nu'}
\left(\frac{\Delta\phi}{2\pi}\right)\ ,
\end{eqnarray}
where $\delta\equiv\nu/\nu'=1/\gamma(1-\beta\cos\theta)\approx
2\gamma/(1+\gamma^2\theta^2)$ is the Doppler factor, which is
given by
\begin{equation}\label{delta}
\delta= \left\{\matrix{(R_L/\eta cT)x & x<x_{\rm dec}\cr\cr
\frac{2(4-k)\gamma_L x^{(k-3)/2}}{4-k+x^{k-4}-1} & x>x_{\rm dec}
}\right.\ .
\end{equation}

For $T<T_{\rm dec}$ we have $R_L=2\eta^2cT<2\eta^2cT_{\rm
dec}=R_{\rm dec}$, and
$L'_{\nu/\delta}(R)=L'_{\nu/2\eta}(R_L)x^a[\delta/2\eta]^{-b}
=L'_{\nu/2\eta}(R_L)x^{a-b}$, and therefore
\begin{equation}\label{F_nu1}
F_\nu(T)=\frac{2\eta L'_{\nu/2\eta}[R_L(T)]}{4\pi
D^2}\int_0^1dx\,x^{1+a-b}\frac{\Delta\phi(x)}{2\pi}= \frac{2\eta
L'_{\nu/2\eta}(R_{\rm dec})}{4\pi D^2}\left(\frac{T}{T_{\rm
dec}}\right)^{a}\int_0^1dx\,x^{1+a-b}\frac{\Delta\phi(x)}{2\pi}\ .
\end{equation}
As long as the outer edge of the ring is not seen,
$\Delta\phi/2\pi=1$, and we have a result similar to the spherical
case, where $F_\nu(T)\propto T^{a}$. 

For $T>T_{\rm dec}$ we have $R_L(T)>R_{\rm dec}$, and the integral
in Eq. \ref{F_nu} naturally divides into two terms, corresponding
to $R<R_{\rm dec}$ and $R_{\rm dec}<R<R_L$, respectively.
Therefore
\begin{eqnarray}
F_\nu(T)&=&\frac{2\eta L'_{\nu/2\eta}(R_{\rm dec})}{4\pi
D^2}\left\{\left(\frac{T}{T_{\rm
dec}}\right)^{b-2} \int_0^1dy\,y^{1+a-b}\frac{\Delta\phi(y)}{2\pi}
\right.\nonumber \\
\label{F_nu2} & & \quad\quad +\left. x_{\rm dec}^{-a+(1-b)(3-k)/2}
\int_{x_{\rm dec}}^1
dx\,x^{a-2+(3-b)(5-k)/2}\left[\frac{1+(3-k)x^{4-k}}{(4-k)}\right]^{b-2}
\frac{\Delta\phi(x)}{2\pi} \right\}\ ,\quad\quad
\end{eqnarray}
where $y=R/R_{\rm dec}$, and it should be noted that the values of
$a$ are generally different in the two integrals. The values of $b$
may also change in the middle of each of the two integrals, in which
case this would require to divide the range of integration
accordingly, and use the appropriate value of $b$ in each sub-range.
From Eq. \ref{F_nu2} it can be seen that the second term dominates at
$T\gg T_{\rm dec}$, which in the spherical case (where
$\Delta\phi/2\pi=1$) implies $F_\nu\propto x_{\rm
dec}^{-a+(1-b)(3-k)/2}\propto T^{-[-a+(1-b)(3-k)/2]/(4-k)}$, since
$x_{\rm dec}\propto 1/R_L(T)\propto T^{-1/(4-k)}$.

Please note that $2\eta L'_{\nu/2\eta}(R_{\rm dec})=L_\nu(R_{\rm
dec},\theta=0)$, which means that the coefficient in front of the
integrals in Eqs.  \ref{F_nu1} and \ref{F_nu2} is
approximately\footnote{This is only approximate since there are
significant contributions to the observed flux from
$\theta\lesssim\gamma^{-1}$ from which the Doppler factor is somewhat
lower than its value exactly along the line of sight at $\theta=0$.}
$F_\nu(T_{\rm dec})$, for a viewing angle within the jet, which can be
calculated from the corresponding values of the peak flux and break
frequencies,
\begin{eqnarray}
F_{\rm\nu,max}(T_{\rm dec}) &=&\left\{\matrix{
7.8(1+z)\epsilon_{B,-2}^{1/2}n_0^{1/2}E_{\rm iso,52}d_{L28}^{-2}\;{\rm
  mJy}
& \quad & k=0\ ,\cr\cr
2.4\times 10^5(1+z)\epsilon_{B,-2}^{1/2}A_*^{3/2}\eta_{2.5}^2 d_{L28}^{-2}\;{\rm
  mJy} &\quad & k=2\ ,
}\right.
\\ \nonumber
\\ \nonumber
\\
\nu_m(T_{\rm dec}) &=& \left\{\matrix{
4.1\times 10^{18}g^2(1+z)^{-1}\epsilon_{B,-2}^{1/2}\epsilon_{e,-1}^2 n_0^{1/2}
\eta_{2.5}^4\;{\rm Hz}
& \quad & k=0\ ,\cr\cr
1.3\times 10^{23}g^2(1+z)^{-1}\epsilon_{B,-2}^{1/2}\epsilon_{e,-1}^2
A_*^{3/2}E_{\rm iso,52}^{-1}\eta_{2.5}^6 d_{L28}^{-2}\;{\rm
  mJy} &\quad & k=2\ ,
}\right.
\\ \nonumber
\\ \nonumber
\\
\nu_c(T_{\rm dec}) &=& \left\{\matrix{
1.6\times 10^{17}(1+z)^{-1}(1+Y)^{-2}\epsilon_{B,-2}^{-3/2}n_0^{-5/6}
E_{\rm iso,52}^{-2/3}\eta_{2.5}^{4/3}\;{\rm Hz}
& \quad & k=0\ ,\cr\cr
1.1\times
10^{10}(1+z)^{-1}(1+Y)^{-2}\epsilon_{B,-2}^{-3/2}A_*^{-5/2}
E_{\rm iso,52}\eta_{2.5}^{-2}\;{\rm
  Hz} &\quad & k=2\ ,
}\right. 
\end{eqnarray}
where $g=3(p-2)/(p-1)$, $Y$ is the Compton y-parameter, and
$\epsilon_e=0.1\epsilon_{e,-1}$ ($\epsilon_B=0.01\epsilon_{B,-1}$) is
the fraction of the internal energy behind the shock in relativistic
electrons (the magnetic field).

\section{Results}
\label{sec:res}

\subsection{A Two Component Jet}
\label{sec:2comp}

The two component jet model has been suggested as an explanation for
sharp rebrightening features in the afterglow light curves of XRF
030723 \citep{Huang04} and GRB 030329 \citep{Berger03}.  For XRF
030723, \citet{Huang04} suggested that our line of sight is slightly
outside the wide jet, so that the beaming cone of the radiation from
the wide jet expands enough to include the line of sight early on,
while that of the narrow jet does so at a significantly later time,
causing a bump in the light curve\footnote{We note that in this
scenario, the true energy of the narrow jet must be larger than that
of the wide jet in order to produce a bump in the afterglow light
curve \citep{PKG05}.} which might account for the sharp bump seen in
the optical afterglow light curve of XRF 030723 at $T\sim 15\;$days.
In Fig. \ref{030723} we show the light curve calculated using the
model from \S \ref{sec:uniform_ring} with the parameters of the best
fit model from \citet{Huang04}, which clearly shows that the resulting
bump in the light curve is very smooth due to the integration over the
surface of equal arrival time of photons to the observer.  Therefore,
it cannot produce the very sharp rise in the flux at the onset of the
observed rebrightening in the optical afterglow of XRF 030723
\citep{Fynbo04}. An alternative explanation for this bump in the light
curve is a contribution from an underlying supernova component, which
naturally produces the red colors that were observed during the bump
\citep{Fynbo04} and could also potentially produce a sharp enough rise
to the bump \citep{Tominaga04}.

For GRB 030329, \citet{Berger03} suggested that the sharp bump in the
optical afterglow light at $T\sim 1.5\;$days \citep{Lipkin04} is due
to a two component jet model where our line of sight is within the
narrow jet and the bump in the light curve occurs at the deceleration
time of the wide jet, $T_{\rm dec,w}$.  In Fig. \ref{030329} we show
the optical light curve for our model from \S \ref{sec:uniform_ring}
using parameters similar to those used by \citet{Berger03}. Despite
the fact that our model assumes an abrupt hydrodynamic transition at
the deceleration time, between the early coasting phase and the
subsequent self similar deceleration phase, the rise to the bump in
the light curve is not sharp enough to match the observations. If a
more gradual hydrodynamic transition at $T_{\rm dec,w}$ is assumed, as
is shown in Fig. \ref{030329_nle} using model 1 of \citet{GK03}, this
produces a much smoother bump in the light curve, which is in a much
stronger contrast with the observed sharp bump. A much more likely
explanation for the bump in the optical light curve of GRB 030329,
which can also account for the subsequent bumps in the following days
and for the duration of these bumps, is refreshed shocks
\citep{GNP03}. Thus, we conclude that both a jet viewed off-axis
becoming visible and the deceleration of a jet viewed on-axis produce
smooth bumps in the afterglow light curves and cannot account for very
sharp features.

\subsection{A Ring Shaped Jet}
\label{sec:ring}

\begin{figure}
\includegraphics[width=16cm]{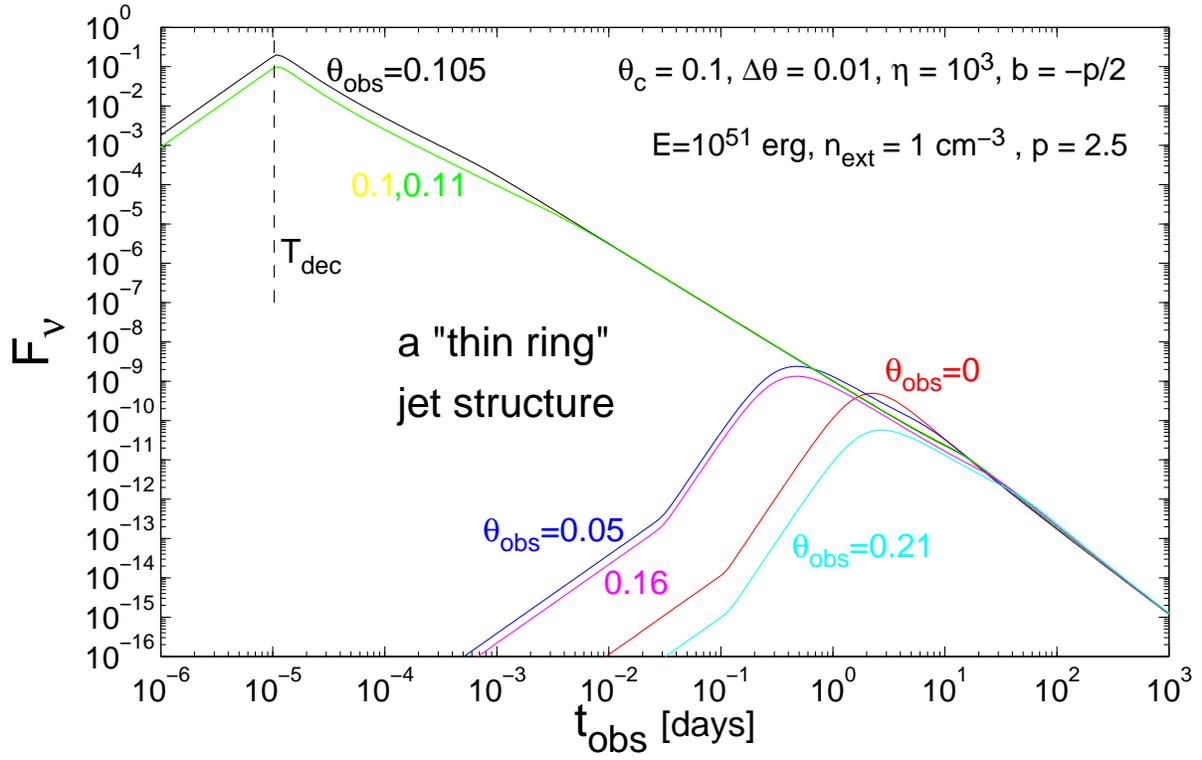}
\caption{Light curves for a jet with a structure of a thin uniform
  ring, using the model from \S \ref{sec:uniform_ring}. The vertical
  dashed line indicates the deceleration time, $T_{\rm dec}$.}
\label{thin_ring}
\end{figure}

Fig. \ref{thin_ring} shows the light curves for a jet with an angular
profile of a thin uniform ring, using the model from \S
\ref{sec:uniform_ring}. The jet occupies
$\theta_c<\theta<\theta_c+\Delta\theta$, where in the example shown in
Fig. \ref{thin_ring}, $\theta_c=0.1$, $\Delta\theta=0.01$ and
$\eta=10^3$. For viewing angles within the jet itself,
$\theta_c\le\theta_{\rm obs}\le\theta_c+\Delta\theta$, the light
curves have a rather small dependence on the exact value of the
viewing angle, $\theta_{\rm obs}$. In fact, the light curves for lines
of sight at the inner edge ($\theta_{\rm obs}=0.1$; yellow line) and
outer edge ($\theta_{\rm obs}=0.11$; green line) of the ring, are
practically one on top of the other (making it hard to see the yellow
line). The deceleration time occurs early on. The ``jet break'' in the
light curve breaks up into two separate and smaller steepening
epochs. The first steepening of the light curve occurs at $T_{j1}$,
when both edges of the ring become visible, i.e. when
$\gamma\Delta\theta\sim 1$ for a line of sight near the inner or outer
edge of the ring, and when $\gamma\Delta\theta\sim 2$ for a line of
sight midway across the width of the ring ($\theta_{\rm
obs}=\theta_c+\Delta\theta/2$). After $T_{j1}$ the light curves from
all of the viewing angles within the jet itself
($\theta_c\le\theta_{\rm obs}\le\theta_c+\Delta\theta$) become
practically indistinguishable, while before $T_{j1}$ there are small
differences, up to a factor of 2. The second steepening of the light
curve occurs at $T_{j2}$, when all of the jet becomes visible,
i.e. when $\gamma\theta_c\sim 1/2$. The light curves from $\theta_{\rm
obs}=0$ and $\theta_{\rm obs}=2\theta_c+\Delta\theta\approx 2\theta_c$
join those for $\theta_c\le\theta_{\rm obs}\le\theta_c+\Delta\theta$
at a slightly earlier time when $\gamma\theta_c\sim 1$, while the
light curves from $\theta_{\rm obs}=\theta_c/2$ or $\theta_{\rm
obs}=1.5\theta_c+\Delta\theta$ join in earlier on, when
$\gamma\theta_c\sim 2$. The fact that the jet break is divided into
two distinct steepening epoch in the light curve, with half of the
total steepening at each epoch, implies that this model cannot
reproduce the large steepening at a single jet break time in the light
curve as is observed in GRB afterglows. Therefore, this jet
structure does not work well for GRB jets.

\begin{figure}
\includegraphics[width=16cm]{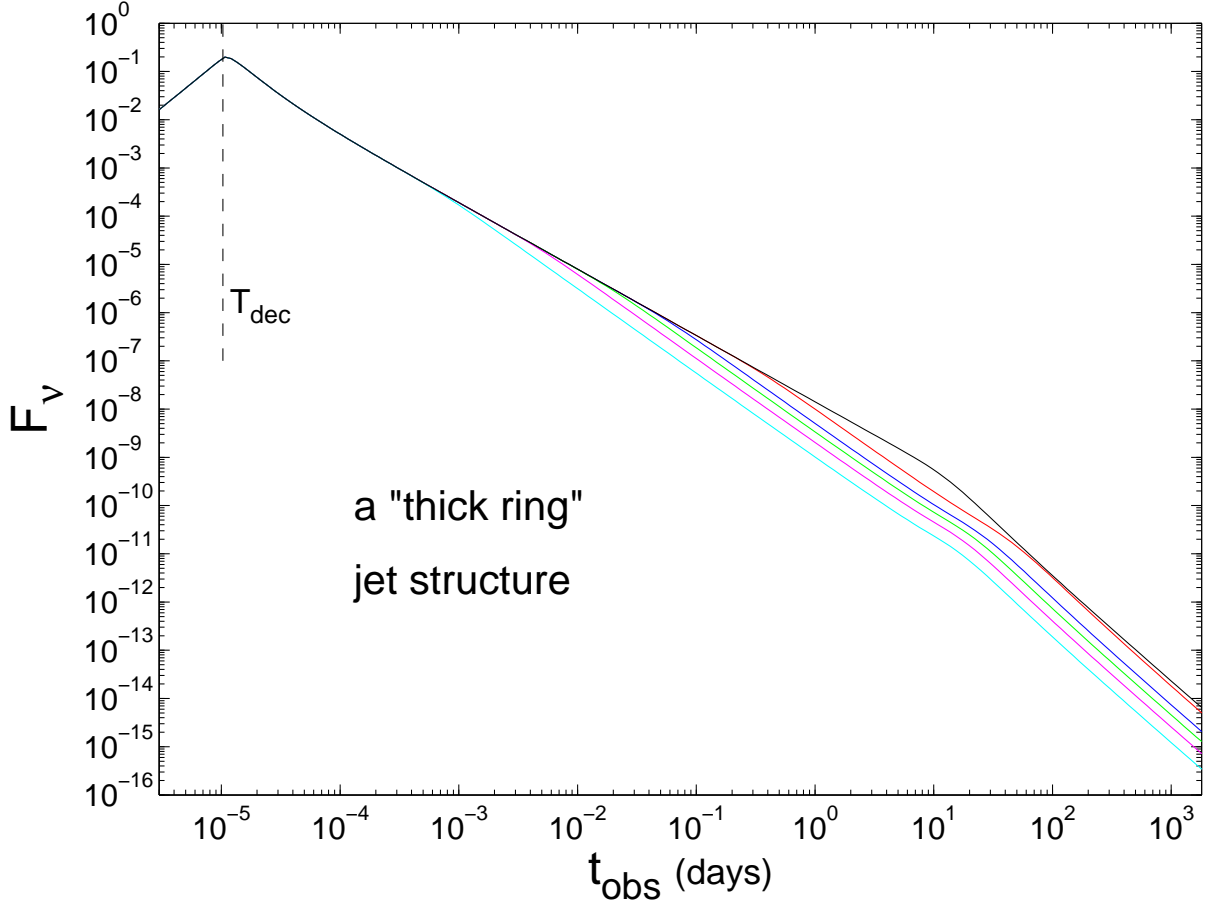}
\caption{Light curves for a jet with an angular structure of a ring
  for various fractional widths, viewed from within the jet.  The
  upper line is for a uniform jet viewed from along its symmetry axis
  ($\theta_c=\theta_{\rm obs}=0$, $\Delta\theta=0.2$) and is included
  for comparison, while the other lines are for a ring shaped jet with
  $\theta_c=0.1$ and $\theta_c/\Delta\theta=1,\,2,\,3,\,5,\,10$, from
  top to bottom, viewed from $\theta_{\rm
  obs}=\theta_c+\Delta\theta/2$. The light curves are calculated using
  the model from \S \ref{sec:uniform_ring}, and for a constant energy
  per solid angle within the jet (corresponding to the same value as
  in Fig. \ref{thin_ring}). We also use $k=0$, $n_0=1$, $b=-p/2$ and
  $p=2.5$.}
\label{thick_ring}
\end{figure}

There is, however, a theoretical motivation for a ``thick ring'' jet
structure \citep{EL03,LE04,EL04}, where $\theta_c/\Delta\theta\sim
2-3$. In Fig. \ref{thick_ring} we show the light curves for a line of
sight within the jet itself ($\theta_{\rm
obs}=\theta_c+\Delta\theta/2$) for different values of the ratio
$\theta_c/\Delta\theta$ which correspond to different fractional
widths of the ring. We keep $\theta_c=0.1$ and the energy per solid
angle within the jet constant, while varying $\theta_c/\Delta\theta$.
It can be seen that, as expected, $T_{j1}$ is smaller for larger
values of $\theta_c/\Delta\theta$ which correspond to a narrower ring,
while $T_{j2}$ remains roughly constant. We note that even for
$\theta_c/\Delta\theta$ as low as 1, the two steepening epochs in the
light curve, $T_{j1}<T_{j2}$, are still quite distinct and separated
by $\sim 1-2$ orders of magnitude in time. For comparison, we also
show the light curve for a uniform jet viewed on-axis
($\theta_c=\theta_{\rm obs}=0$, $\Delta\theta=0.2$), which produces a
single sharp jet break in the afterglow light curve, similar to those
observed in GRB afterglows.  

\begin{figure}
\includegraphics[width=16cm]{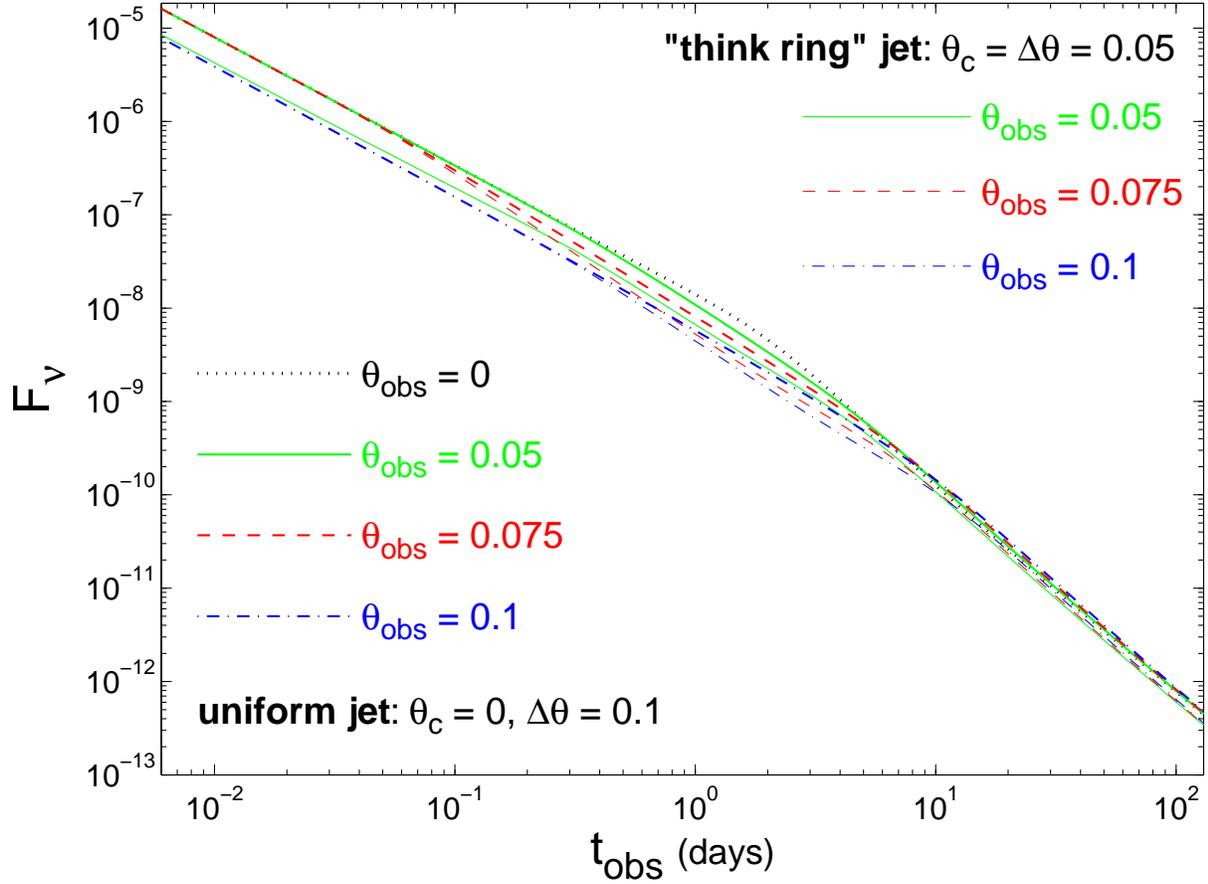}
\caption{Light curves for a jet with an angular structure of a thick
  ring compared to those for a uniform conical jet, for different
  viewing angles. The light curve are calculated using the model from
  \S \ref{sec:uniform_ring} with $k=0$, $n_0=1$, $b=-p/2$ and
  $p=2.5$. The same energy per solid angle is used for the two jet
  structures.}
\label{thick_ring_vs_uniform}
\end{figure}

In Fig. \ref{thick_ring_vs_uniform} we show the light curves for a
``thick ring'' jet ($\theta_c=\Delta\theta=0.05$) together with those
for a uniform conical jet ($\theta_c=0$, $\Delta\theta=0.1$) with the
same outer angle and the same energy per solid angle, so that the
``thick ring'' jet is obtained from the uniform conical jet by taking
out its inner half (in terms of $\theta$). I can be seen that the
light curves for a uniform jet show a sharper jet break for viewing
angle near the jet symmetry axis, and somewhat smoother jet breaks for
viewing angles closer to the outer edge of the jet. This result is
similar to that from numerical simulations \citep{Granot01}. By
comparing the light curves for the two jet structures viewed from the
same viewing angle we can see that those for the ``thick ring'' jet
produce a somewhat lower flux as they are missing the contribution
from the central part of the jet. The ``thick ring'' jet also produce
a somewhat less pronounced jet break compared to a uniform jet.
However, for $\Delta\theta\gtrsim\theta_c$ the differences in the
light curves compared to those for a uniform conical jet might not be
large enough to easily distinguish between these two jet structures
using the observed afterglow light curves. Furthermore, for
$\Delta\theta\sim\theta_c$, if there is relativistic lateral expansion
of the jet in its own rest frame, this might help bring $T_{j1}$ and
$T_{j2}$ closer together, making the light curves closer to the
observations. For $\Delta\theta<\theta_c$, however, the effects of
lateral spreading should be rather small. We therefore conclude that a
ring shaped jet requires a very thick ring, with
$\Delta\theta\gtrsim\theta_c$, in order to reproduce the observed
afterglow light curves, while a jet in the shape of a thinner ring
does not produce afterglow light curves with a sharp enough jet break
to match afterglow observations.

\subsection{A Fan Shaped Jet}
\label{sec:fan}

An interesting jet structure which resembles a fan can arise due to a
magneto-centrifugally launched wind that is driven by the newly formed
proto-neutron star during the supernova explosion \citep{Thompson04}.  If
this wind is concentrated within a narrow angle $\theta_0$ around the
rotational equator and somehow makes it out of the star while still
highly relativistic, this could create a GRB outflow within
$|\tilde{\theta}|<\theta_0$ where $\tilde{\theta}=\theta-\pi/2$ is the
angle from the rotational equator (i.e. the latitude). The fraction of
the total solid angle that is occupied by such a jet is
$f_b=\sin\tilde{\theta}_j\approx\tilde{\theta}_j$. This corresponds to
a ring shaped jet with the parameters $\theta_c=\pi/2-\Delta\theta/2$
and $\Delta\theta=2\theta_0$, using the notations from \S
\ref{sec:uniform_ring}.

\begin{figure}
\includegraphics[width=16cm]{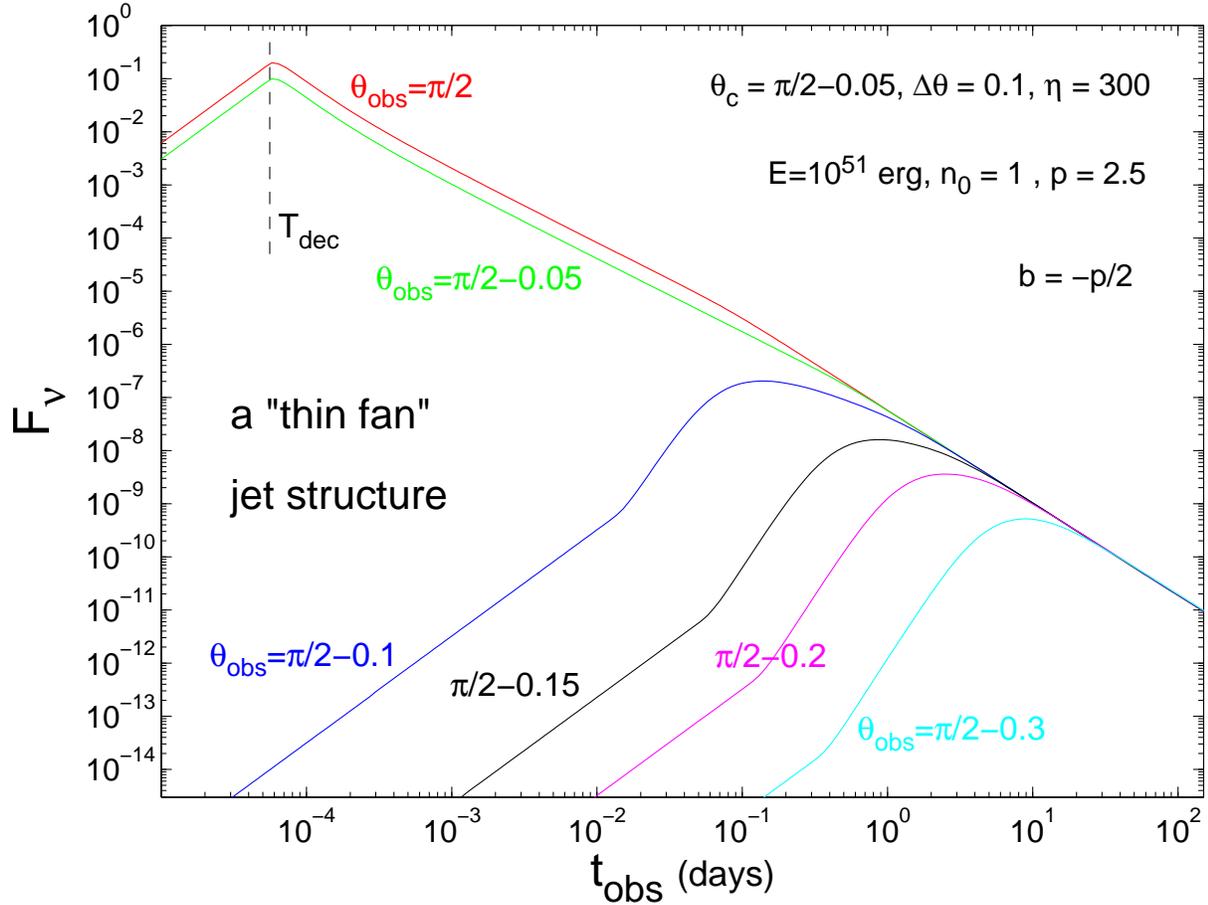}
\caption{Light curves for a jet with an angular structure of a thin
  fan, with an opening angle of $\Delta\theta=0.1$ centered on
  $\theta=\pi/2$
  (i.e. $\theta_c=\pi/2-\Delta\theta/2=\pi/2-0.05$). The light curves
  are calculated using the model from \S \ref{sec:uniform_ring}.}
\label{fan}
\end{figure}

If there is no lateral expansion, then the steepening during the jet
break is by a factor of $\sim\gamma\theta_0\propto T^{-(3-k)/2(4-k)}$
corresponding to $\Delta\alpha=(3-k)/2(4-k)$ which is $3/8$ for $k=0$
and $1/4$ for $k=2$.  This is much shallower than observed in the jet
breaks of GRB afterglows, and exactly half as steep (in terms of
$\Delta\alpha$) as the jet break for a conical uniform jet. This is
demonstrated in Fig. \ref{fan} which shows light curves for this jet
structure that were calculated using the model from \S
\ref{sec:uniform_ring}. As for the ring shaped jet, the light curves
for viewing angles within the jet are similar to each other, differing
by up to a factor of 2 before the jet break time, and practically
identical after the jet break time. In contrast to the ring shaped
jet, there is only one jet break time in the light curve, with at most
half the steepening compared to a uniform conical jet.  This may be
understood as in the limit $\theta_c\sim\pi/2$, $T_{j1}$ becomes
similar to the time of the non-relativistic transition, and therefore
does not produce a distinct break in the light curve, so that the only
one epoch of steepening in the light curve remains, at $T_{j1}$, when
the edges of the fan shaped jet become visible (i.e. when
$\gamma\Delta\theta\sim 1$ for a line of sight at the edge of the jet
and when $\gamma\Delta\theta\sim 2$ for a line of sight at the center
of the jet).  The light curves for lines of sight outside of the jet
join those for lines of sight inside the jet when the beaming cone of
the radiation from the jet (which extend out to an angle of $\sim
1/\gamma$ form the edges of the jet) reaches the line of sight. 

For a relativistic expansion in the local rest frame we
have $\tilde{\theta}_j\approx\max(\tilde{\theta}_0,\gamma^{-1})$ so
that at $T>T_j$, $\tilde{\theta}_j\approx\gamma^{-1}$ and
\begin{equation}
E\approx\frac{4\pi}{(3-k)}Ac^2 R^{3-k}\gamma^2\tilde{\theta}\approx
\frac{4\pi}{(3-k)}Ac^2 R^{3-k}\gamma \ ,
\end{equation}
implying 
\begin{equation}
\gamma\propto R^{k-3}\propto T^{-(3-k)/(7-2k)}\ .
\end{equation}
This behavior is intermediate between the spherical case,
$\gamma\propto T^{-(3-k)/2(4-k)}$, and the case of a narrow conical
jet that expands sideways relativistically in its own rest frame,
$\gamma\propto T^{-1/2}$. The temporal decay index at $T>T_j$ is also
intermediate: $F_\nu\propto T^{(12-5k-24p+7kp)/4(7-2k)}$ for
$\nu_m<\nu<\nu_c$ and $F_\nu\propto T^{(8-2k-24p+7kp)/4(7-2k)}$ for
$\nu>\max(\nu_c,\nu_m)$.  In both cases $\Delta\alpha$ is reduced by a
factor of $(7-2k)/(3-k)$ compared to a uniform conical jet, i.e. the
jet break less than half as steep. This result is also valid for the
first jet break (at $T_{j1}$) for a a jet in the shape narrow ring,
that was discussed in \S \ref{sec:ring}. 

Finally, we have so far assumed that the fan shaped jet occupies all of
the range of azimuthal angle $\varphi$. If it occupies a smaller
range, $\Delta\varphi<2\pi$, then as long as $\Delta\varphi\gtrsim 1$
the second jet break will overlap with the non-relativistic transition
and would not produce a distinct steepening of the light curve.  For
$\Delta\theta\ll\Delta\varphi\ll 1$, however, there will be two
distinct light curves, the first the same as described above and a
second jet break when the edges of the jet in the $\varphi$ direction
become visible (when $\gamma\Delta\varphi\sim 1$ for a line of sight
at the edge of the jet in the $\varphi$ direction, and at
$\Delta\varphi\sim 2$ for a line of sight at the center of the jet in
the $\varphi$ direction).

\section{Conclusions}
\label{sec:conc}

In \S\S \ref{sec:thin_sphere} and \ref{sec:uniform_ring} we have
developed a semi-analytic formalism for calculating the the afterglow
light curves for a jet with an angular profile of a uniform ring.  The
final expression for the observed flux is the sum of two one
dimensional integrals which are trivial to evaluate
numerically. Despite its simplicity, this model includes integration
over the surface of equal arrival time of photons to the observer,
thus producing realistic results when this is indeed the dominant
effect in smoothing out sharp features in the afterglow light curve.

\begin{figure}
\includegraphics[width=15cm]{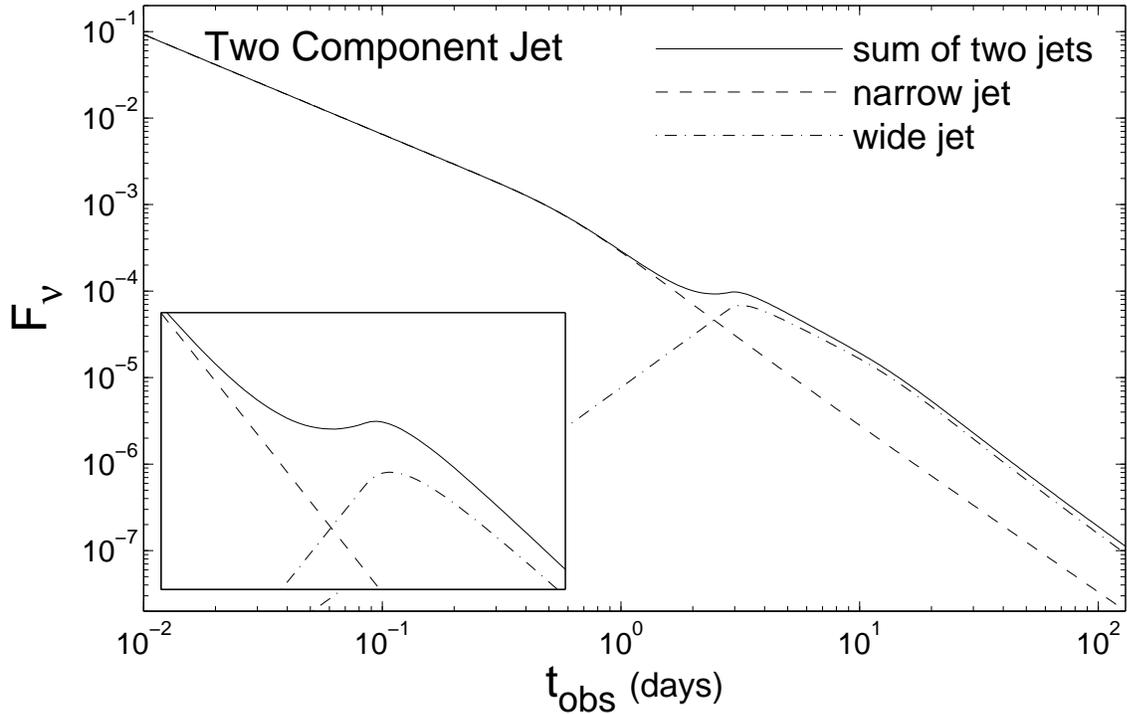}
\caption{Optical light curve for a two component jet model were the
  physical parameters of the jet were taken to be similar to those
  used by \citet{Berger03} in order to account for the multi frequency
  afterglow light curves of GRB 030329, namely $\theta_n=0.09$,
  $\theta_w=0.3$, $\theta_{\rm obs}=0$, $E_{n,{\rm iso}}=1.2\times
  10^{52}\;$erg, $E_{w,{\rm iso}}=5.6\times 10^{51}\;$erg,
  $\eta_w=6.5$, $k=0$, $n_0=1.8$, $p=2.2$ and $b=-p/2$. The inset
  shows a close up of the bump in the light curve that occurs near the
  deceleration time of the wide jet, as its emission starts to
  dominate the observed flux.}
\label{030329}
\end{figure}

\begin{figure}
\includegraphics[width=15cm]{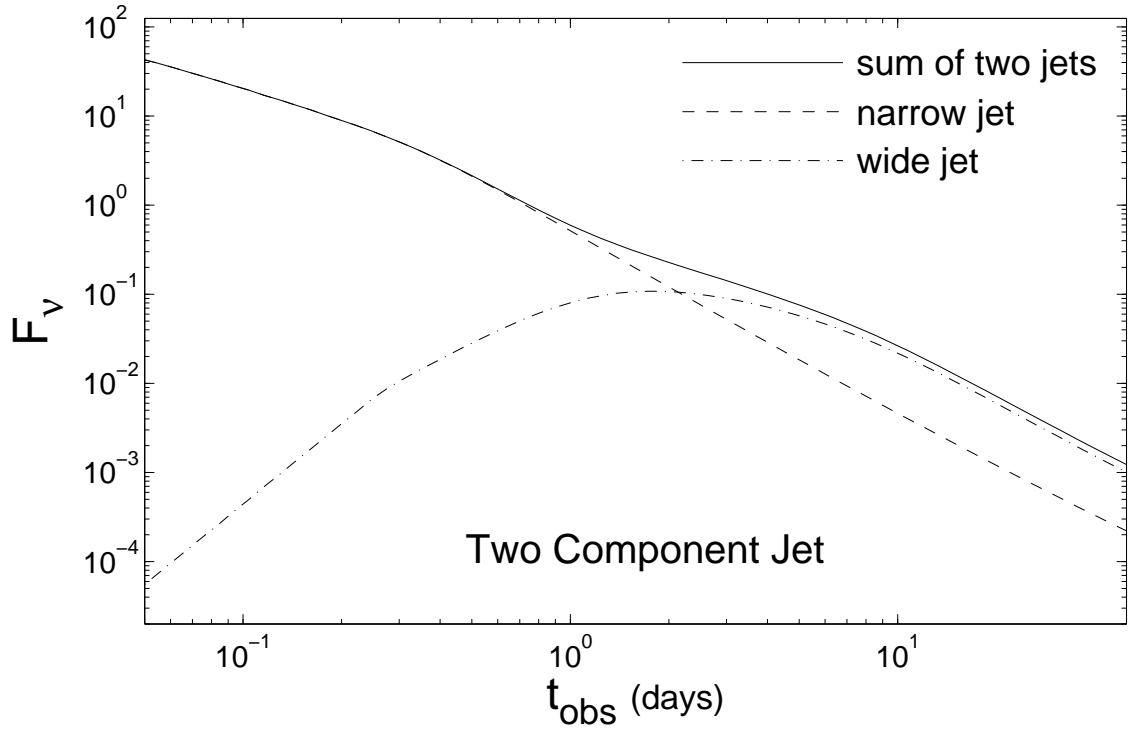}
\caption{Similar to Fig. \ref{030329} but using a different numerical
  code \citep[model 1 of][]{GK03} which features a more gradual
  dynamical transition at the deceleration time, resulting in a much
  smoother bump in the light curve near $T_{\rm dec,w}$.}
\label{030329_nle}
\end{figure}

The price of the simple expressions for the observed flux is simple
assumptions on the jet dynamics, namely no lateral expansion and an
abrupt transition at the deceleration time, $T_{\rm dec}$.  This
results in a relatively sharp peak in the light curve at $T_{\rm
dec}$, while a more realistic model for the dynamics with a smoother
dynamical transition at $T_{\rm dec}$ would produce a smoother peak in
the light curve (compare Figs. \ref{030329} and \ref{030329_nle}).
Nevertheless, when used with care, this simple formalism may serve as
a powerful tool. It may also be generalized so that it could be
applicable to other jet structures, which were not considered in this
work, or to include a calculation of the polarization assuming some
local configuration of the magnetic field (e.g., a field tangled
within the plane of the shock which is identified with the thin
emitting shell).

\begin{figure}
\includegraphics[width=15cm]{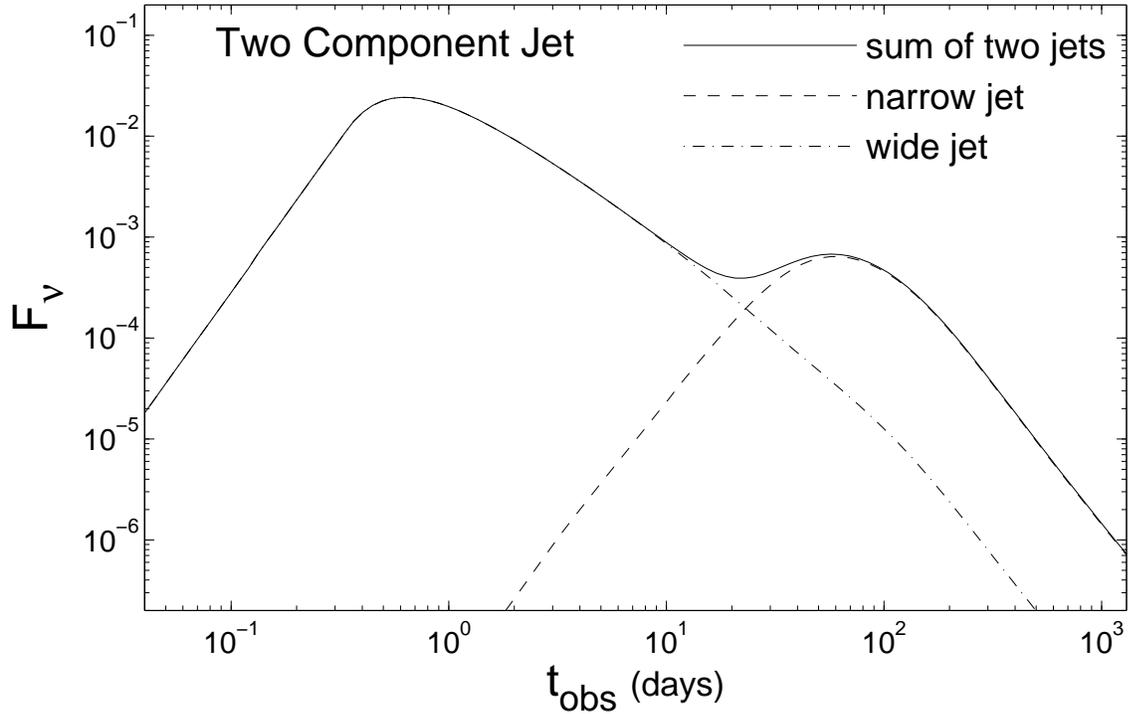}
\caption{Optical light curve for a two component jet model. The
  physical parameters of the jet were taken from the best fit model of
  \citet{Huang04} to the optical light curve of XRF 030723:
  $\theta_n=0.09$, $\theta_w=0.3$, $\theta_{\rm obs}=0.37$, $E_{n,{\rm
  iso}}=3\times 10^{53}\;$erg, $E_{w,{\rm iso}}=10^{52}\;$erg, $k=0$,
  $n_0=1$ and $p=3.2$. The exact values of $\epsilon_e$ and
  $\epsilon_B$ effect the flux normalization but not the shape of the
  light curve, as long as a break frequency does not pass through the
  observed frequency band, as is assumed here since we use a constant
  $b=(1-p)/2$.}
\label{030723}
\end{figure}

In \S \ref{sec:2comp} we have shown that the two-component jet model
cannot produce very sharp features in the afterglow light curve, due
to the deceleration of the wide (or narrow) jet, or when the narrow
(or wide) jet becomes visible at lines of site outside of the jet
aperture, as the beaming cone of the emitted radiation reaches the line
of sight. Therefore, such explanations for the bumps in the optical
light curves of GRB 030329 \citep{Berger03} and XRF 030723
\citep{Huang04}, which both had a sharp rise to the bump, do not work
well (see Figs. \ref{030723}, \ref{030329} and \ref{030329_nle}).

The afterglow light curves for a jet with a ring-like, or ``hollow
cone'' angular profile were calculated in \S \ref{sec:ring}.  We find
that the jet break in the light curve divides into two distinct
steepening episodes, $T_{j1}$ and $T_{j2}$, with roughly (or in our
simple model, exactly) half of the total steepening occurring at each
of these two times. The two times remain distinct even for a
moderately thick ring, and might merge into a single jet break in the
light curve only for a very thick ring, with
$\Delta\theta\gtrsim\theta_c$.

The light curves for a ``fan'' shaped jet were calculated in \S
\ref{sec:fan}, and show a single jet break in the light curve with a
very moderate steepening across the break, which is at most half of
that for a ``standard'' conical uniform jet. The jet break is even
slightly shallower when lateral expansion is taken into account, in
which case it is less than half of the steepening for a conical
uniform jet. Such a shallow jet break cannot account for the large
steepening of the light curves that are observed in GRB afterglows.

\acknowledgments 

I thank T.~A. Thompson, A. K\"onigl, D. Eichler and E. Ramirez-Ruiz
for useful comments.  This research was supported by US Department
of Energy under contract number DE-AC03-76SF00515.

\newcommand{\rb}[1]{\raisebox{1.5ex}[0pt]{#1}}
\begin{deluxetable}{cccc}
%\tabletypesize{\scriptsize} %\rotate
\tabletypesize{\footnotesize} \tablecaption{The Radial Dependence of
  the Local Rest Frame Spetral Luminosity} \tablewidth{0pt} \tablehead{
\colhead{PLS} & \colhead{$b$} & \colhead{$a$ ($R<R_{\rm dec}$)} &
\colhead{$a$ ($R>R_{\rm dec}$)}
}
\startdata
D & $1/3$ & $3-k/2$ & $3-4k/3$
\\ 
E & $1/3$ & $11/3-2k$ & $(5-4k)/3$
\\
F & $-1/2$ & $2-3k/4$ & $(5-2k)/4$
\\
G & $(1-p)/2$ & $3-k(p+5)/4$ & $[15-9p-2k(3-p)]/4$
\\ 
H & $-p/2$ & $2-k(p+2)/4$ & $[14-9p+2k(p-2)]/4$
\enddata \tablecomments{\label{table1}The luminosity in the local rest
  frame of the emitting fluid behind the afterglow shock scales as a
  power law in frequency and in radius, $L'_{\nu'}\propto
  R^a(\nu')^b$, where the power law indexes $a$ and $b$ change between
  the different power law segments (PLSs) of the spectrum. In
  addition, $a$ also changes between $R<R_{\rm dec}$ and $R>R_{dec}$.
  The first column labels the power law segment of the spectrum
  following the notation of \citet{GS02}. The second column provides
  the value of spectral index $b$, while the third and fourth columns
  give the value of $a$ for $R<R_{\rm dec}$ and $R>R_{\rm dec}$,
  respectively.}
%\tablenotetext{\dagger}{}
%\tablenotetext{\dagger\dagger}{}
%\tablenotetext{a}{Sample footnote for table~\ref{tbl-1} that was
%generated with the deluxetable environment}
%\tablenotetext{b}{Another sample footnote for table~\ref{tbl-1}}

\end{deluxetable}

\clearpage

\clearpage

\clearpage

\clearpage

\clearpage

\clearpage

\clearpage

\end{document}